\documentclass[aip,apl,reprint,groupedaddress]{revtex4-1}

\usepackage{graphicx}

\begin{document}

\title{Sparse reconstructions of overlapping three-dimensional point spread functions using overcomplete dictionaries}

\author{Anthony Barsic}
\email[Corresponding author: ]{anthony.barsic@colorado.edu}
\author{Ginni Grover}
\author{Rafael Piestun}
\affiliation{
Department of Electrical, Computer, and Energy Engineering, University of Colorado at Boulder\\
UCB 425, Boulder, Colorado 80309, USA
}

\begin{abstract} When a single molecule is detected in a wide-field microscope, the image approximates the point spread function of the system.  However, as the distribution of molecules becomes denser and their images begin to coincide, existing solutions to determine the number of molecules present and their precise three-dimensional locations can tolerate little to no overlap.  A solution to this problem involving matched optical and digital techniques, as here proposed, is critical to increase the allowable labeling density and to accelerate single-molecule localization microscopy.  \end{abstract}

\maketitle

\noindent In single-molecule localization microscopy \cite{Betzig2006, Rust2006, Hess2006}, sparse sets of emitters are localized by identifying well separated single-molecule images and fitting them to high precision, thereby achieving resolution better than the diffraction limit.  Similar problems appear in many biological and biophysical experiments where two or more molecules need to be resolved or their distance estimated \cite{Weiss2000, Yildiz2004, DeLuca2006}.  Localization precision can be much better than the diffraction limit, depending on the number of photons detected from the emitter and noise conditions \cite{Heisenberg1930}.  Subsequently, photoswitching, photoactivation, and other mechanisms were proposed and developed to overcome the problem of overlapping molecule images in a time sequential form \cite{Betzig2006, Rust2006, Hess2006, Folling2008}.  The trade-off for super-resolution in these methods is a slower acquisition rate---typically, tens of thousands of frames are collected and processed to generate a single super-resolution image.  To ameliorate this problem, researchers have proposed fitting schemes that allow for a few emitters generating overlapping images \cite{Zhu2012, Cox2012, Huang2011, Barsic2013, Babcock2012}.  Unfortunately, except for [\onlinecite{Babcock2012}], which offers a limited increase in density, all methods reported so far are limited to two-dimensional imaging.  Nevertheless, in most cases three-dimensional (3D) information is required for complete understanding of the phenomenon under examination.

To further advance the allowable labeling density, we propose and investigate enhanced methods for solving this problem, namely finding the number and 3D locations of clustered emitters from a single image. The experimental demonstration of the technique in biological samples opens up new opportunities to acquire quantitative information about single molecules and other emitters that remain unresolved in three dimensions with conventional methods. The proposed methods also enable faster acquisition times for 3D single-molecule localization microscopy, which is critical for live-cell super-resolution imaging.

In what follows we emphasize the distinction between the image generated by an emitter, such as a single-molecule, and the point spread function (PSF). While the former depends on the emission pattern of the emitter, noise, sample induced aberrations, and the detector array, the latter is only a function of the optical imaging system.

The key observation behind the methods proposed here is the fact that raw images in single-molecule localization microscopy are a combination of sparse (possibly overlapping) molecule images and noise from different sources.  This raw image can be efficiently represented with a dictionary consisting of the PSFs of transversely and longitudinally shifted point emitters. A dictionary is a set of vectors that spans the space of possible images.

Dictionaries provide alternate representations to the pixel-based image; i.e. a set of coefficients describing the degree to which each dictionary element is present in the image.  Interestingly, a scene that appears dense to our eyes (contains numerous overlapping images) may be sparse in a properly chosen dictionary.  Sparse means the image can be expressed by a number of coefficients K that is significantly smaller than the number of pixels used in the scene N ($K<<N$).  We note that the most efficient representation of a scene with overlapping single-molecule images contains a single coefficient for each emitter in the scene.  Since each coefficient in the solution corresponds to an emitter, they are intrinsically resolved, and the coefficients are easily converted to locations and photon counts.

The method for resolving and localizing 3D clusters of single molecules involves a combination of optical and digital techniques: (a) Imaging the sample with a proper 3D PSF imaging system; (b) Creating a model of the system via experimental measurements, theoretical calculations, or a combination of both; (c) Establishing a dictionary composed of the PSF for different locations of a point source in a dense 3D grid; (d) Solving the estimation problem of determining the coefficients of the dictionary elements that best represent the data. Once the non-zero coefficients are known, the number of molecules and their locations and brightnesses can be determined.

Several techniques can encode depth information onto a two-dimensional image by utilizing a more complex PSF \cite{Kao1994, Ram2007b, Pavani2008}.  Without losing generality, we chose the double-helix (DH) PSF  because of its inherent precision and depth of field advantages \cite{Pavani2008}.  Accordingly, a single emitter in the focal plane generates an image with two horizontally displaced lobes.  The transverse location of the emitter is related to the center of the two lobes, and the axial location is encoded in the orientation of the lobes \cite{Quirin2012}.  For illustration, a few dictionary elements for a DH-PSF system are shown in Fig. 1.  Each dictionary element contains the 2D cross-section of the PSF corresponding to a different discrete emitter location in the full 3D space.  Note that the methods demonstrated here can be applied to any 3D PSF and are not limited to the DH-PSF. Simulations confirm that these methods also work with the astigmatic PSF.

\begin{figure}
\includegraphics[width=6cm]{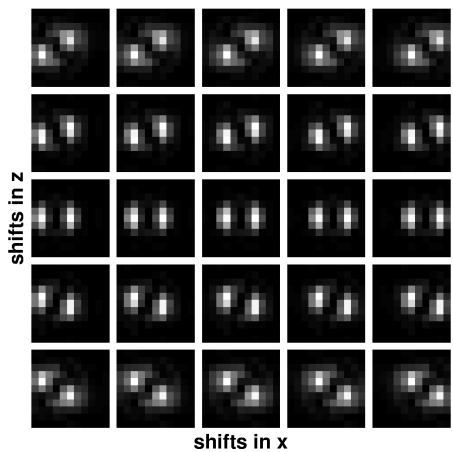}
\caption{Dictionary example. An example of a selection of dictionary elements for a Double-Helix system.}
\end{figure}

The most important design parameter for the dictionaries is the step size between adjacent elements.  Localization precision is limited by step size, so small steps are desirable.  Conversely, smaller steps mean more elements, and hence a more computationally difficult problem.  Typical imaging system designs and desired localization precision necessitate sub-pixel steps.  Furthermore, the dictionary extends along the axial direction.  These two factors mean we require dictionaries that are overcomplete, i.e. there are more elements D in the dictionary than there are pixels per element ($D>N$).

To solve the estimation problem, we investigate two methods that are representative of large classes of solvers.  The first method is Matching Pursuit (MP), which is iterative.  In MP, an iteration consists of projecting the image onto the dictionary, finding and storing the largest coefficient, and subtracting that element from the image \cite{Bergeaud}.  Iterations continue until a stopping criterion is met.  The second method is a direct solver using Convex Optimization (CO) \cite{Boyd2004}.  The reconstruction problem is formulated as a convex problem in which the variable to be optimized is the sparsity of the coefficient vector (quantified as the L1 norm).  This method attempts to arrive at a solution for the significant coefficients in parallel. The method was implemented with CVX \cite{Grant}.

To demonstrate the performance of the two methods, we first present a simulation of a single-molecule experiment containing several overlapping molecule images embedded in noise.  To match our experimental system, the effective pixel sizes in sample space are 160nm.  Three molecules are positioned randomly in a small volume of 3D space to generate the simulated raw images.

For MP, the dictionary step size is 10nm in the transverse dimensions and 15nm in depth, yielding 1.27 million dictionary elements.  If the reconstruction returns the correct elements, the expectation value of the error due to the quantization of the solution space is 5.75nm.  Even with such a large dictionary, one MP solution on an 11-by-11 window can be completed in tenths of a second on a desktop computer.  A simulation containing three emitters and noise is shown in Fig. 2a, and the solution is shown in Fig. 2b.  MP correctly recognizes there are three emitters, and returns locations with an average error of 77nm.

\begin{figure}
\includegraphics[width=8cm]{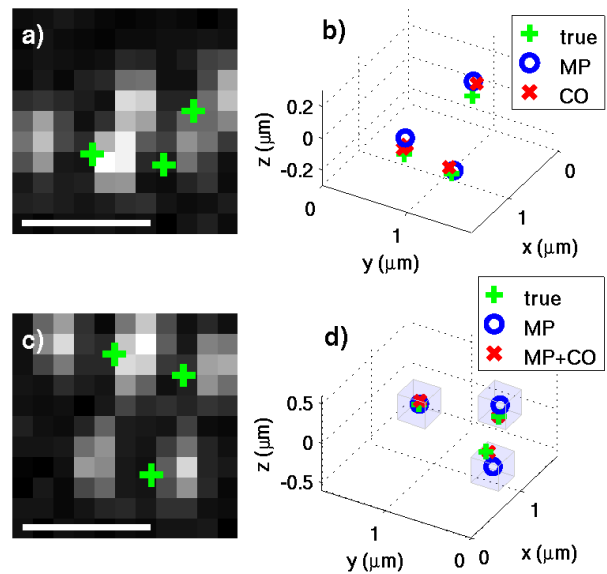}
\caption{Reconstruction of sparse clusters of emitters. (a) and (c) Simulation examples of three emitters imaged with a DH-PSF system. (b) shows the resulting localizations using Matching Pursuit (MP) and Convex Optimization (CO) for example (a). It illustrates the superior localization capability of CO, at the expense of computational complexity.  In (d), the hybrid method (MP+CO) is applied to (b), showing the true locations, the intermediate MP results, and the hybrid localizations. The blue boxes indicate the regions included in the refined dictionary for the final CO step.  The mean localization error for MP was 118 nm while the use of MP+CO resulted in a mean error of 34 nm.  In all images, scale bars are $1 \mu m$.  Signal levels were designed to mimic experimental data (1000-1200 photons per emitter).}
\end{figure}

Due to computational limitations, the dictionary for CO cannot be as large because it is a much more computationally intensive algorithm.  Thus, we select a coarser dictionary with transverse steps of 80nm and axial steps of 120nm (2904 elements, expected error of 46nm).  The result of the CO reconstruction achieves the optimal solution, with an average localization error of 44nm.  However, even with the significantly smaller dictionary, CO still requires an order of magnitude more computation time than MP with a very fine dictionary.  For comparison, MP with this coarse dictionary only requires a few milliseconds.

Monte Carlo simulations show MP is faster, but the results are sub-optimal.  Conversely, the reconstruction obtained with CO is orders of magnitude slower, but the returned locations achieve the limit imposed by the fineness of the dictionary.  To attain a method that is fast, accurate, and precise we developed a hybrid algorithm that takes advantage of the strengths of both methods.  First, MP provides a rough estimate of the number of emitters and their locations.  Since the precise location is not needed at this stage, we use a coarse dictionary.  Next, we perform CO, but this time with a refined dictionary.  Interestingly, because coarse estimates of the locations are already available from the MP method, we can limit the dictionary to only include elements located close to those estimates.  Therefore, we implement a dictionary with the desired fineness while the problem is still computationally tractable.  This technique reduces the size of the fine dictionary by nearly two orders of magnitude, enabling CO to be performed in a reasonable time.

An example of the hybrid method (MP+CO) applied to simulated data is shown in Fig. 2c-d.  From the location plot in (d), it is clear that the results from MP, while imprecise, are still providing reliable estimates.  This level of accuracy is sufficient to enable an automated size reduction of the finest dictionary (expected quantization error of 5.75nm for a step size of 10 (15) nm in transverse (axial) directions).  The localization precision achieved with this methods is in-line with current state-of-the-art single-molecule localization microscopy, while being able to handle higher densities than previously achievable for 3D localization.

An example of an experimental scene (raw data) of a dense molecule cluster is shown in Fig. 3.  The image was acquired using a SPINDLE system \cite{Grover2012} incorporating a DH-PSF and stochastic optical reconstruction using photoswitchable dyes \cite{Rust2006}.  The scene in Fig. 3a presents a complex arrangement of single molecule images.  MP+CO is able to resolve and localize three emitters in this scene, despite the overlap of the DH-PSF lobes.  Such a scene would be rejected from typical localization algorithms, and none of the emitters would contribute to the final image.

\begin{figure}
\includegraphics[width=8cm]{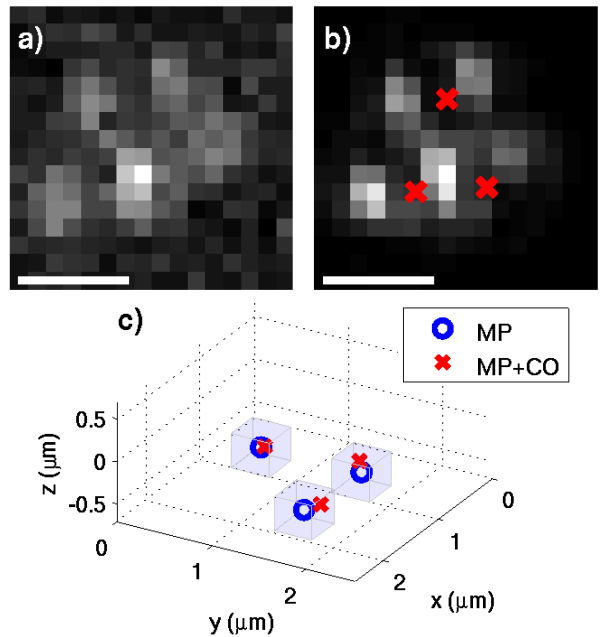}
\caption{Experimental demonstration of 3D super-resolution and super-localization from overlapping single-molecule images. (a) an example of raw data of overlapping molecule images using a DH-PSF system. The image in (b) shows the reconstructed image using MP+CO.  (c) shows the estimated locations of the molecules using the MP and hybrid methods.  The regions used for the refined dictionary are marked with blue cubic boxes.  In all images, scale bars are $1 \mu m$.}
\end{figure}

\begin{figure}
\includegraphics[width=8cm]{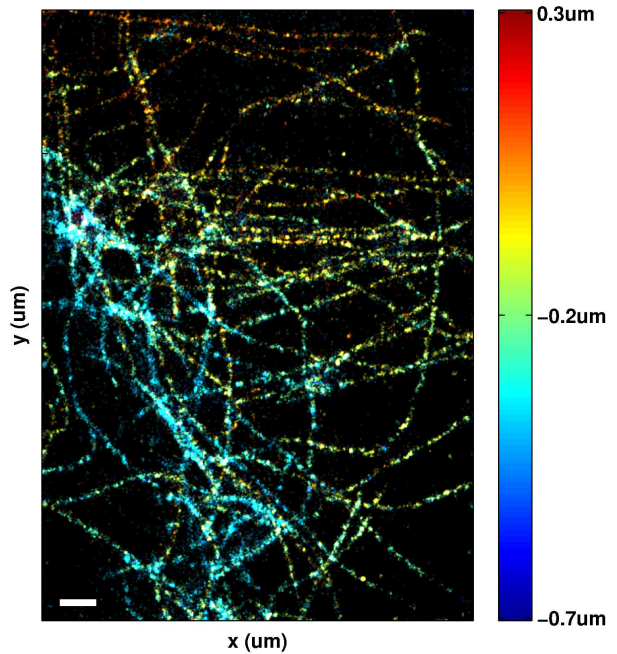}
\caption{Large-scale experimental implementation. This 3D super-resolution image of labeled tubulin in PtK1 cells demonstrates that the method can be scaled to wide field of view localization-based super-resolution imaging. Scale bar: $1 \mu m$.}
\end{figure}

An example application in a large-scale super-resolution image is shown in Fig. 4.  The sample is composed of PtK1 cells (Rat Kangaroo Epithelial cells) in which tubulin is labeled with Alexa-647 and Alexa-488 dyes.  This image was compiled from more than 30,000 frames, and it clearly demonstrates 3D super-resolution capabilities.  In the standard fluorescence image of this scene, many of the individual microtubules either cannot be resolved, or are out of focus due to the limited depth of focus of the standard PSF of high-NA objectives.

It is interesting to point out the origin of the super-resolution capability, given that the raw images are limited by diffraction and it is well known that deconvolution techniques have had only limited success \cite{McNally1999}. First, it is essential to use a structured 3D PSF to retrieve 3D information and to form a 3D dictionary. Second, the fundamental assumption of sparsity, i.e. only up to a handful of emitters are located within the PSF area, provides the required prior knowledge to achieve effective resolution and localization. Hence, it is the combination of 3D optical techniques and prior knowledge that ultimately enable super-resolution/localization.

In conclusion, the method presented here has the capability to three-dimensionally resolve denser clusters of molecules (or other emitters) from a single image than previously possible, while maintaining the high 3D localization precision.  Therefore, higher labeling density and significant decreases in data collection time for super-resolution microscopy experiments are now possible.

\section*{Acknowledgements}
This work was supported in part by the National Science Foundation award DGE-0801680, an Integrative Graduate Education and Research Traineeship program in Computational Optical Sensing and Imaging.  We thankfully acknowledge Jennifer and Keith DeLuca for providing the samples, and Eyal Niv for helpful discussions.


\begin{thebibliography}{10}
\expandafter\ifx\csname url\endcsname\relax
  \def\url#1{\texttt{#1}}\fi
\expandafter\ifx\csname urlprefix\endcsname\relax\def\urlprefix{URL }\fi
\providecommand{\bibinfo}[2]{#2}
\providecommand{\eprint}[2][]{\url{#2}}

\bibitem{Betzig2006}
\bibinfo{author}{Betzig, E.} \emph{et~al.}
\newblock \bibinfo{title}{{Imaging intracellular fluorescent proteins at
  nanometer resolution}}.
\newblock \emph{\bibinfo{journal}{Science}} \textbf{\bibinfo{volume}{313}},
  \bibinfo{pages}{1642--5} (\bibinfo{year}{2006}).

\bibitem{Rust2006}
\bibinfo{author}{Rust, M.~J.}, \bibinfo{author}{Bates, M.} \&
  \bibinfo{author}{Zhuang, X.}
\newblock \bibinfo{title}{{Sub-diffraction-limit imaging by stochastic optical
  reconstruction microscopy (STORM)}}.
\newblock \emph{\bibinfo{journal}{Nature Methods}}
  \textbf{\bibinfo{volume}{3}}, \bibinfo{pages}{793--795}
  (\bibinfo{year}{2006}).

\bibitem{Hess2006}
\bibinfo{author}{Hess, S.~T.}, \bibinfo{author}{Girirajan, T. P.~K.} \&
  \bibinfo{author}{Mason, M.~D.}
\newblock \bibinfo{title}{{Ultra-high resolution imaging by fluorescence
  photoactivation localization microscopy.}}
\newblock \emph{\bibinfo{journal}{Biophysical Journal}}
  \textbf{\bibinfo{volume}{91}}, \bibinfo{pages}{4258--72}
  (\bibinfo{year}{2006}).

\bibitem{Weiss2000}
\bibinfo{author}{Weiss, S.}
\newblock \bibinfo{title}{{Measuring conformational dynamics of biomolecules by
  single molecule fluorescence spectroscopy.}}
\newblock \emph{\bibinfo{journal}{Nature structural biology}}
  \textbf{\bibinfo{volume}{7}}, \bibinfo{pages}{724--9} (\bibinfo{year}{2000}).

\bibitem{Yildiz2004}
\bibinfo{author}{Yildiz, A.}, \bibinfo{author}{Tomishige, M.},
  \bibinfo{author}{Vale, R.~D.} \& \bibinfo{author}{Selvin, P.~R.}
\newblock \bibinfo{title}{{Kinesin walks hand-over-hand.}}
\newblock \emph{\bibinfo{journal}{Science (New York, N.Y.)}}
  \textbf{\bibinfo{volume}{303}}, \bibinfo{pages}{676--8}
  (\bibinfo{year}{2004}).

\bibitem{DeLuca2006}
\bibinfo{author}{DeLuca, J.~G.} \emph{et~al.}
\newblock \bibinfo{title}{{Kinetochore microtubule dynamics and attachment
  stability are regulated by Hec1.}}
\newblock \emph{\bibinfo{journal}{Cell}} \textbf{\bibinfo{volume}{127}},
  \bibinfo{pages}{969--82} (\bibinfo{year}{2006}).

\bibitem{Heisenberg1930}
\bibinfo{author}{Heisenberg, W.}
\newblock \emph{\bibinfo{title}{{The Physical Principles of the Quantum
  Theory}}} (\bibinfo{publisher}{University of Chicago Press, Chicago},
  \bibinfo{year}{1930}).

\bibitem{Folling2008}
\bibinfo{author}{F\"{o}lling, J.} \emph{et~al.}
\newblock \bibinfo{title}{{Fluorescence nanoscopy by ground-state depletion and
  single-molecule return}}.
\newblock \emph{\bibinfo{journal}{Nature Methods}}
  \textbf{\bibinfo{volume}{5}}, \bibinfo{pages}{943--945}
  (\bibinfo{year}{2008}).

\bibitem{Zhu2012}
\bibinfo{author}{Zhu, L.}, \bibinfo{author}{Zhang, W.},
  \bibinfo{author}{Elnatan, D.} \& \bibinfo{author}{Huang, B.}
\newblock \bibinfo{title}{{Faster STORM using compressed sensing.}}
\newblock \emph{\bibinfo{journal}{Nature Methods}}
  \textbf{\bibinfo{volume}{9}}, \bibinfo{pages}{721--3} (\bibinfo{year}{2012}).

\bibitem{Cox2012}
\bibinfo{author}{Cox, S.} \emph{et~al.}
\newblock \bibinfo{title}{{Bayesian localization microscopy reveals nanoscale
  podosome dynamics}}.
\newblock \emph{\bibinfo{journal}{Nature Methods}}
  \textbf{\bibinfo{volume}{9}}, \bibinfo{pages}{195--200}
  (\bibinfo{year}{2012}).

\bibitem{Huang2011}
\bibinfo{author}{Huang, F.}, \bibinfo{author}{Schwartz, S.~L.},
  \bibinfo{author}{Byars, J.~M.} \& \bibinfo{author}{Lidke, K.~A.}
\newblock \bibinfo{title}{{Simultaneous multiple-emitter fitting for single
  molecule super-resolution imaging}}.
\newblock \emph{\bibinfo{journal}{Biomedical Optics Express}}
  \textbf{\bibinfo{volume}{2}}, \bibinfo{pages}{1377--93}
  (\bibinfo{year}{2011}).

\bibitem{Barsic2013}
\bibinfo{author}{Barsic, A.} \& \bibinfo{author}{Piestun, R.}
\newblock \bibinfo{title}{{Super-resolution of dense nanoscale emitters beyond
  the diffraction limit using spatial and temporal information}}.
\newblock \emph{\bibinfo{journal}{Applied Physics Letters}}
  \textbf{\bibinfo{volume}{102}}, \bibinfo{pages}{231103}
  (\bibinfo{year}{2013}).

\bibitem{Babcock2012}
\bibinfo{author}{Babcock, H.}, \bibinfo{author}{Sigal, Y.~M.} \&
  \bibinfo{author}{Zhuang, X.}
\newblock \bibinfo{title}{{A high-density 3D localization algorithm for
  stochastic optical reconstruction microscopy}}.
\newblock \emph{\bibinfo{journal}{Optical Nanoscopy}}
  \textbf{\bibinfo{volume}{1}}, \bibinfo{pages}{6} (\bibinfo{year}{2012}).

\bibitem{Kao1994}
\bibinfo{author}{Kao, H.~P.} \& \bibinfo{author}{Verkman, a.~S.}
\newblock \bibinfo{title}{{Tracking of single fluorescent particles in three
  dimensions: use of cylindrical optics to encode particle position.}}
\newblock \emph{\bibinfo{journal}{Biophysical journal}}
  \textbf{\bibinfo{volume}{67}}, \bibinfo{pages}{1291--300}
  (\bibinfo{year}{1994}).

\bibitem{Ram2007b}
\bibinfo{author}{Ram, S.}, \bibinfo{author}{Chao, J.},
  \bibinfo{author}{Prabhat, P.}, \bibinfo{author}{Ward, E.~S.} \&
  \bibinfo{author}{Ober, R.~J.}
\newblock \bibinfo{title}{{A novel approach to determining the
  three-dimensional location of microscopic objects with applications to 3D
  particle tracking}}.
\newblock \emph{\bibinfo{journal}{Proceedings of SPIE}}
  \textbf{\bibinfo{volume}{6443}}, \bibinfo{pages}{64430D--64430D--7}
  (\bibinfo{year}{2007}).

\bibitem{Pavani2008}
\bibinfo{author}{Pavani, S. R.~P.} \& \bibinfo{author}{Piestun, R.}
\newblock \bibinfo{title}{{High-efficiency rotating point spread functions.}}
\newblock \emph{\bibinfo{journal}{Optics express}}
  \textbf{\bibinfo{volume}{16}}, \bibinfo{pages}{3484--9}
  (\bibinfo{year}{2008}).

\bibitem{Quirin2012}
\bibinfo{author}{Quirin, S.}, \bibinfo{author}{Pavani, S. R.~P.} \&
  \bibinfo{author}{Piestun, R.}
\newblock \bibinfo{title}{{Optimal 3D single-molecule localization for
  superresolution microscopy with aberrations and engineered point spread
  functions.}}
\newblock \emph{\bibinfo{journal}{Proceedings of the National Academy of
  Sciences of the United States of America}} \textbf{\bibinfo{volume}{109}},
  \bibinfo{pages}{675--9} (\bibinfo{year}{2012}).

\bibitem{Bergeaud}
\bibinfo{author}{Bergeaud, F.} \& \bibinfo{author}{Mallat, S.}
\newblock \bibinfo{title}{{Matching pursuit of images}}.
\newblock In \emph{\bibinfo{booktitle}{Image Processing, 1995. Proceedings.,
  International Conference on}}, vol.~\bibinfo{volume}{1},
  \bibinfo{pages}{53--56} (\bibinfo{year}{1995}).

\bibitem{Boyd2004}
\bibinfo{author}{Boyd, S.} \& \bibinfo{author}{Vandenberghe, L.}
\newblock \emph{\bibinfo{title}{{Convex Optimization}}}
  (\bibinfo{publisher}{Cambridge University Press}, \bibinfo{year}{2004}).

\bibitem{Grant}
\bibinfo{author}{Grant, M.} \& \bibinfo{author}{Boyd, S.}
\newblock \bibinfo{title}{{CVX: Matlab software for disciplined convex
  programming, version 2.0 beta}}. \bibinfo{url}{\url{http://cvxr.com/cvx}} (\bibinfo{year}{2012}).

\bibitem{Grover2012}
\bibinfo{author}{Grover, G.}, \bibinfo{author}{DeLuca, K.},
  \bibinfo{author}{Quirin, S.}, \bibinfo{author}{DeLuca, J.} \&
  \bibinfo{author}{Piestun, R.}
\newblock \bibinfo{title}{{Super-resolution photon-efficient imaging by
  nanometric double-helix point spread function localization of emitters
  (SPINDLE).}}
\newblock \emph{\bibinfo{journal}{Optics express}}
  \textbf{\bibinfo{volume}{20}}, \bibinfo{pages}{26681--95}
  (\bibinfo{year}{2012}).

\bibitem{McNally1999}
\bibinfo{author}{McNally, J.~G.}, \bibinfo{author}{Karpova, T.},
  \bibinfo{author}{Cooper, J.} \& \bibinfo{author}{Conchello, J.~a.}
\newblock \bibinfo{title}{{Three-dimensional imaging by deconvolution
  microscopy.}}
\newblock \emph{\bibinfo{journal}{Methods (San Diego, Calif.)}}
  \textbf{\bibinfo{volume}{19}}, \bibinfo{pages}{373--85}
  (\bibinfo{year}{1999}).

\end{thebibliography}

\end{document}